\documentclass{article}

\usepackage{arxiv}

\usepackage[utf8]{inputenc} 
\usepackage[T1]{fontenc}    
\usepackage{hyperref}       
\usepackage{url}            
\usepackage{booktabs}       
\usepackage{amsfonts}       
\usepackage{nicefrac}       
\usepackage{microtype}      
\usepackage{lipsum}		
\usepackage{graphicx}
\usepackage{doi}
\usepackage{amsmath}
\usepackage{authblk}
\usepackage[title]{appendix}
\usepackage[sort&compress,numbers]{natbib}

\title{On the closure of 2D flamelet theory in non-orthogonal composition spaces}

\author[ ~,1]{Hernan Olguin\thanks{Corresponding author\\ Email address: hernan.olguin@usm.cl}}
\author[2]{Pascale Domingo}
\author[2]{Luc Vervisch}
\author[3]{Arne Scholtissek}
\author[3]{Christian Hasse}

\affil[1]{Department of Mechanical Engineering, Universidad Técnica Federico Santa María,\linebreak Avenida España 1680, Valparaíso, Chile\linebreak}
\affil[2]{CORIA-CNRS, INSA de Rouen, Technop\^ole du Madrillet, BP 8, Saint-\'Etienne-du-Rouvray 76801, France\linebreak}
\affil[3]{Institute for Simulation of reactive Thermo-Fluid Systems, TU Darmstadt,\linebreak Otto-Berndt-Stra{\ss}e 2, 64287 Darmstadt, Germany}

\date{}

\begin{document}
\maketitle

\begin{abstract}
Formal relations between scalar dissipation rates and curvatures are developed in the context of non-orthogonal two-dimensional composition spaces. These show that it is possible to achieve general closure of flamelet theory in terms of strain rates only, but that this requires the consideration of at least three different parameters (both in non-orthogonal and orthogonal spaces).
\end{abstract}

\vspace{15mm}

\section{Introduction}

During the last decades, several two-dimensional extensions of Peters' non-premixed flamelet equations have been proposed in the literature (see for example~\cite{Hasse05,Domingo08,Knudsen09,Nguyen10,Knudsen12,Mueller20,Scholtissek20,Novoselov21,Olguin25}). While these typically introduce convenient simplifications, written in their most general form they would inevitably contain the same unclosed variables: i) two normal scalar dissipation rates, ii) two curvatures and, for non-orthogonal composition spaces, iii) a cross scalar dissipation rate. Since these are all geometric quantities, formal relations between them must exist, a fact that has been so far actively exploited in the context of orthogonal coordinates only~\cite{Olguin24}.

This work provides general mathematical relations between normal scalar dissipation rates, curvatures and the cross scalar dissipation rate, which enable closure of 2D non-orthogonal flamelet equations in terms of three strain rates only. Additionally, the applicability and implications of the formalism are discussed.

\section{Preliminaries and closure requirements}
\label{section2}
We introduce a general 2D non-orthogonal coordinate system built in terms of the mixture fraction, $Z$ and a second generic variable (for instance a progress variable), $\beta$, which are governed by 
\begin{align}
    \frac{D Z}{D t} = \Gamma_{Z}~\text{      }~\text{   and     }~\text{       }~\frac{D \beta}{D t} = \Gamma_{\beta},
\label{generaleq}
\end{align}
respectively, where $D(\cdot)/Dt$ is a material derivative and the $\Gamma$-terms can comprise diffusion and other effects of interest such as chemical reactions. 

Based on $Z$ and $\beta$, we define now the scalar gradients, $g_{Z} = \lvert \nabla Z \rvert$ and $g_{\beta} = \lvert \nabla \beta \rvert$, and the unit vectors $\mathbf{n}_Z = \nabla Z / \lvert \nabla Z \rvert$ and $\mathbf{n}_{\beta} = \nabla \beta / \lvert \nabla \beta \rvert$, which are required to construct: i) The classical normal scalar dissipation rates, $\chi_Z = 2 D g_Z^2$ and $\chi_{\beta} = 2 D g_{\beta}^2$, ii) the curvatures, $\kappa_Z = - \nabla \cdot \mathbf{n}_Z$ and $\kappa_{\beta} = - \nabla \cdot \mathbf{n}_{\beta}$, and iii) the cross scalar dissipation rate, $\chi_{Z\beta} = 2 D g_Z g_{\beta} \Theta$, where $\Theta = (\mathbf{n}_Z \cdot \mathbf{n}_{\beta})$ quantifies the degree of alignment between the coordinates $Z$ and $\beta$.

For the closure of $g_Z$, a transport equation can be obtained from the $Z$-equation after applying the operator $\mathbf{n}_Z\cdot \nabla (\cdot)$. This leads to~\cite{Olguin232}
\begin{align}
    \frac{D g_Z}{D t} = a_Z g_Z + \frac{\partial \Gamma_Z}{\partial n_Z},
\end{align}
where $a_Z = - \mathbf{n}_Z \cdot \nabla \mathbf{u} \cdot \mathbf{n}_Z$ is a normal strain rate and $\partial (\cdot)/\partial n_Z = \mathbf{n}_Z \cdot \nabla (\cdot)$ is a directional derivative. 

Introducing now a change of coordinates from physical, ($\mathbf{x}, t$), to composition space, ($Z, \beta, \tau$), with $\tau$ denoting a time-like variable, the transformation rules for temporal and spatial derivatives yield
\begin{equation}
    \frac{\partial (\cdot)}{\partial t} = \frac{\partial (\cdot)}{\partial \tau} + \frac{\partial (\cdot)}{\partial Z} \frac{\partial Z}{\partial t} + \frac{\partial (\cdot)}{\partial \beta} \frac{\partial \beta}{\partial t}
    \label{trans1}
\end{equation}
and
\begin{equation}
\nabla (\cdot) = \frac{\partial (\cdot)}{\partial Z} \nabla Z + \frac{\partial (\cdot)}{\partial \beta} \nabla \beta,
\label{trans2}
\end{equation}
respectively. With these, the $g_Z$-equation can be transformed as~\cite{Olguin232,Olguin25}
\begin{equation}
    \frac{\partial g_Z}{\partial \tau} = - \Gamma_Z \frac{\partial g_Z}{\partial Z} - \Gamma_{\beta} \frac{\partial g_Z}{\partial \beta} + a_Z g_Z + \frac{\partial \Gamma_Z}{\partial n_Z}.
\label{gs-flamelet}
\end{equation}
While curvature and $\Theta$ do not explicitly appear in this form of the $g_Z$-equation, it is highlighted that these quantities are comprised in the $\Gamma$-term. In particular, for the mixture fraction we have
\begin{align}
    \Gamma_Z & = \frac{1}{\rho} \nabla \cdot \left(\rho D \nabla Z\right) \nonumber \\ & = \frac{g_Z}{\rho} \frac{\partial (\rho D g_Z)}{\partial Z} + \frac{g_{\beta}}{\rho} \frac{\partial (\rho D g_Z)}{\partial \beta} \Theta  - D g_Z \kappa_Z,
\end{align}
where the product rule and the flamelet transformation have been used. 

Since this procedure can be replicated to obtain equivalent equations for $g_{\beta}$, the current specific closure requirements for 2D flamelet equations in non-orthogonal composition spaces consist in finding appropriate relations for $\Theta$, $\kappa_Z$ and $\kappa_{\beta}$. In the following sections, closure of these quantities is discussed in detail.

\section{Curvature equations}
\label{section3}
We proceed now to derive explicit equations for $\kappa_Z$ and $\kappa_{\beta}$ in terms of $g_Z$, $g_{\beta}$ and $\Theta$. For this, we introduce an auxiliary unit vector, $\mathbf{n}_{\perp}$, which satisfies $(\mathbf{n}_Z \cdot \mathbf{n}_{\perp}) = 0 $. With this, we can express $\mathbf{n}_{\beta}$ as
\begin{equation}
    \mathbf{n}_\beta = \Theta~\mathbf{n}_Z + \sqrt{1-\Theta^2} \mathbf{n}_{\perp},
\end{equation}
which, together with the definition of curvature and the product rule, leads to
\begin{align}
    \kappa_{\beta} & = - \nabla \cdot \mathbf{n}_{\beta} \nonumber \\ & = - \nabla \cdot \left( \Theta~\mathbf{n}_Z \right) - \nabla \cdot \left(\sqrt{1-\Theta^2}~\mathbf{n}_{\perp} \right) \nonumber \\ & = \Theta \kappa_Z - \frac{ \partial \Theta}{\partial n_Z} + \sqrt{1-\Theta^2} \kappa_{\perp} -  \frac{\partial \left( \sqrt{1-\Theta^2} \right)}{\partial n_{\perp}},
    \label{kappabeta}
\end{align}
The second term at the right hand side (RHS) of this equation can be re-expressed as 
\begin{align}
    -\frac{\partial \Theta}{\partial n_Z} & = - \mathbf{n}_Z \cdot \left( \frac{\partial \Theta}{\partial Z} g_Z \mathbf{n}_Z + \frac{\partial \Theta}{\partial \beta} g_{\beta} \mathbf{n}_{\beta}\right) \nonumber \\ & = - \frac{\partial \Theta}{\partial Z} g_Z - \frac{\partial \Theta}{\partial \beta} g_{\beta} \Theta,
    \label{term2}
\end{align}
while, making use of the following relation~\cite{Olguin24}
\begin{align}
    \kappa_{\perp} = \frac{1}{g_Z} \mathbf{n}_{\perp} \cdot \nabla g_Z,
\end{align}
the third term can be transformed as follows
\begin{align}
    \sqrt{1-\Theta^2} \kappa_{\perp} & = \sqrt{1-\Theta^2} \frac{1}{g_{Z}} \mathbf{n}_{\perp} \cdot \nabla g_Z \nonumber \\ & = \sqrt{1-\Theta^2} \frac{1}{g_{Z}} \mathbf{n}_{\perp} \cdot \left( \frac{\partial g_Z}{\partial Z} \nabla Z + \frac{\partial g_Z}{\partial \beta} \nabla \beta \right) \nonumber \\ & = \left( 1 - \Theta^2 \right) \frac{g_{\beta}}{g_Z} \frac{\partial g_Z}{\partial \beta}.
    \label{term3}
\end{align}
Finally, the fourth term can be worked out as
\begin{align}
- \frac{\partial \left(\sqrt{1-\Theta^2} \right)}{\partial n_{\perp}}  & = \frac{\Theta}{\sqrt{1-\Theta^2}} \mathbf{n}_{\perp} \cdot \nabla \Theta \nonumber \\ & = \frac{\Theta}{\sqrt{1-\Theta^2}} \left( \frac{\partial \Theta}{\partial \beta}  g_{\beta} \left(\mathbf{n}_{\perp} \cdot \mathbf{n}_{\beta}\right)\right)\nonumber \\&= \Theta \frac{\partial \Theta}{\partial \beta}g_{\beta}.
\label{term4}
\end{align}
Inserting Eqs.~(\ref{term2}), (\ref{term3}) and (\ref{term4}) in Eq.(\ref{kappabeta}), we obtain
\begin{equation}
    \kappa_{\beta} =  \Theta \kappa_Z + \left( 1 - \Theta^2 \right) \frac{g_{\beta}}{g_Z} \frac{\partial g_Z}{\partial \beta} - \frac{\partial \Theta}{\partial Z} g_Z.
\label{kappabetafinal}
\end{equation}
Of course, the symmetry between $Z$ and $\beta$ allows deriving an analogous expression for $\kappa_Z$ replicating the procedure presented here, which leads to
\begin{align}
    \kappa_Z = \Theta \kappa_{\beta} + \left( 1 - \Theta^2 \right) \frac{g_Z}{g_{\beta}} \frac{\partial g_{\beta}}{\partial Z} - \frac{\partial \Theta}{\partial \beta} g_{\beta}.
\label{kappazetafinal}
\end{align}
Simultaneously solving Eqs.~(\ref{kappabetafinal}) and (\ref{kappazetafinal}), we obtain
\begin{align}
    \kappa_{Z}  = \frac{g_Z}{g_{\beta}} \frac{\partial g_{\beta}}{\partial Z} + \Theta \frac{g_{\beta}}{g_Z} \frac{\partial g_Z}{\partial \beta}   - \frac{1}{\left(1-\Theta^2 \right)} \left( \Theta \frac{\partial \Theta}{\partial Z} g_Z + \frac{\partial \Theta}{\partial \beta} g_{\beta} \right)
\label{kappa_Z}
\end{align}
and
\begin{align}
    \kappa_{\beta}  = \frac{g_{\beta}}{g_{Z}} \frac{\partial g_{Z}}{\partial \beta} + \Theta \frac{g_{Z}}{g_{\beta}} \frac{\partial g_{\beta}}{\partial Z}  - \frac{1}{\left(1-\Theta^2 \right)} \left( \Theta \frac{\partial \Theta}{\partial \beta} g_{\beta} + \frac{\partial \Theta}{\partial Z} g_{Z} \right),
\label{kappa_beta} 
\end{align}
respectively. In the next section, the closure of $\Theta$ is addressed.

\section{Cross-scalar dissipation rate equation}
\label{section4}
We note that there are two different (and equivalent) ways of expressing the material derivative of the product $(\nabla Z \cdot \nabla \beta) = g_Z g_{\beta} \Theta$. In particular, we have
\begin{align}
    \frac{D \left( \nabla Z \cdot \nabla \beta \right)}{D t}  = g_Z g_{\beta} \frac{D \Theta}{D t}  + \Theta \left( g_Z \frac{D g_{\beta}}{D t} + g_{\beta} \frac{D g_Z}{D t} \right)
    \label{product1}
\end{align}
and
\begin{align}
    \frac{D (\nabla Z \cdot \nabla \beta)}{D t} = \nabla \beta \cdot \frac{D \nabla Z}{D t} + \nabla Z \cdot \frac{D \nabla \beta}{D t}.
    \label{product2}
\end{align}

Applying the operators $\nabla \beta \cdot \nabla (\cdot)$ and $\nabla Z \cdot \nabla (\cdot)$ to the $Z$ and $\beta$-equations, we obtain
\begin{align}
    \nabla \beta \cdot \frac{D \nabla Z}{D t} = - g_{\beta} \left(\mathbf{n}_{\beta} \cdot \nabla \mathbf{u} \cdot  \mathbf{n}_{Z}\right) g_Z + g_{\beta} \frac{\partial \Gamma_{Z}}{\partial n_{\beta}}
    \label{relation1}
\end{align}
and
\begin{align}
    \nabla Z \cdot \frac{D \nabla \beta}{D t} = - g_Z \left(\mathbf{n}_{Z} \cdot \nabla \mathbf{u} \cdot  \mathbf{n}_{\beta}\right) g_{\beta} + g_Z \frac{\partial \Gamma_{\beta}}{\partial n_Z},
    \label{relation2}
\end{align}
respectively. Combining Eqs.~(\ref{product1}) and (\ref{product2}), and making use of Eqs.~(\ref{relation1})~and~(\ref{relation2}), we obtain
\begin{align}
    \frac{D \Theta}{D t}   =  - \Theta \left( a_Z + a_{\beta} + \frac{1}{g_{\beta}} \frac{\partial \Gamma_{\beta}}{\partial n_{\beta}}  + \frac{1}{g_{Z}} \frac{\partial \Gamma_Z}{\partial n_Z} \right)   +  \left( a_{Z\beta} + \frac{1}{g_{\beta}} \frac{\partial \Gamma_{\beta}}{\partial n_Z} + \frac{1}{g_{Z}} \frac{\partial \Gamma_Z}{\partial n_{\beta}} \right)
    \label{theta1}
\end{align}
where a third (cross) strain rate,
\begin{align}
    a_{Z \beta} = - \mathbf{n}_Z \cdot \left( \nabla \mathbf{u} + \nabla \mathbf{u}^T \right) \cdot \mathbf{n}_{\beta}, 
\end{align}
is included.
Applying the flamelet transformations, Eqs.~(\ref{trans1}) and (\ref{trans2}), and the definition of the directional derivative for non-orthogonal spaces, Eq.~(\ref{theta1}) can be rewritten as
\begin{align}
    \frac{\partial \Theta}{\partial \tau}  = -  \Gamma_{Z} \frac{\partial \Theta}{\partial Z} - \Gamma_{\beta} \frac{\partial \Theta}{\partial \beta}  + a_{Z\beta} - \Theta (a_Z + a_{\beta}) + \left(g_Z^2 \frac{\partial \Gamma_{\beta}}{\partial Z} + g_{\beta}^2 \frac{\partial \Gamma_Z}{\partial \beta}\right) \frac{(1-\Theta^2)}{g_Z g_{\beta}},
    \label{theta_final}
\end{align}
which represents the final equation required to close 2D sets of flamelet equations in terms of three strain rates only. 

\section{Discussion}
\label{section5}

The derived relations for curvatures and the cross scalar dissipation rate, Eqs.~(\ref{kappa_Z}), (\ref{kappa_beta}) and (\ref{theta_final}), complement the already available $g$-equations~\cite{Olguin25}, Eq.~(\ref{gs-flamelet}) and its analogous for $g_{\beta}$, providing formal closure of 2D flamelet theory in terms of three strain rates only. However, the strong degree of coupling between these expressions can complicate their proper implementation and will require the development of appropriate solution strategies for its systematic use. Additionally, the cross strain rate, $a_{Z\beta}$, appears as a new parameter in flamelet theory, which requires special attention and further study before a complete adoption.

The current formalism also has important implications for similar approaches derived in the context of orthogonal composition spaces. As expected, the relations developed for the curvature, Eqs.~(\ref{kappa_Z}) and (\ref{kappa_beta}), reduce to expressions previously reported for orthogonal spaces when $\Theta = 0$~\cite{Olguin24}. In particular, we have
\begin{equation}
    \kappa_{Z} = \frac{g_Z}{g_{\beta}} \frac{\partial g_{\beta}}{\partial Z}~\text{      }~\text{   and     }~\text{       }~\kappa_{\beta} = \frac{g_{\beta}}{g_{Z}} \frac{\partial g_{Z}}{\partial \beta}. 
\end{equation}
However, it must be noted that Eq. (\ref{theta_final}) does not vanish for this particular limit. Instead, the following relation is obtained
\begin{equation}
0 = a_{Z\beta} + \left(g_Z^2 \frac{\partial \Gamma_{\beta}}{\partial Z} + g_{\beta}^2 \frac{\partial \Gamma_Z}{\partial \beta}\right) \frac{1}{g_Z g_{\beta}},
\end{equation}
which can be interpreted as a necessary condition to ensure orthogonality in time. It is important to highlight that this condition has not been identified in previous works addressing orthogonal formulations in terms of two strain rates only~\cite{Olguin24}, and it directly introduces the need of prescribing a third parameter for that formulation too.

\bibliography{references}
\end{document}